\documentclass{bmcart}

\pdfoutput=1
\usepackage[T1]{fontenc}
\usepackage[english]{babel}
\usepackage{float}
\usepackage[utf8]{inputenc}
\usepackage{pslatex}   % for times new roman font
\usepackage{amsmath}
\usepackage{siunitx}

\newcommand{\Acirc}{A_\mathrm{circ}}

\newcommand{\Fscat}{F_\mathrm{scat}}

\newcommand{\Fopt}{F_\mathrm{opt}}

\newcommand{\Pcirc}{P_\mathrm{circ}}
\newcommand{\Pinc}{P_\mathrm{inc}}
\newcommand{\Vpd}{V_\mathrm{PD}}
\newcommand{\Icirc}{I_\mathrm{circ}}
\newcommand{\Iinc}{I_\mathrm{inc}}
\newcommand{\Irefl}{I_\mathrm{refl}}

\usepackage{babel}
\usepackage{graphicx}

\startlocaldefs
\endlocaldefs

\begin{document}

%%% Start of article front matter
\begin{frontmatter}

\begin{fmbox}
\dochead{Research}


\title{Optical excitation of atomic force microscopy cantilever
  for accurate spectroscopic measurements}

%%%%%%%%%%%%%%%%%%%%%%%%%%%%%%%%%%%%%%%%%%%%%%
%%                                          %%
%% Enter the authors here                   %%
%%                                          %%
%% Specify information, if available,       %%
%% in the form:                             %%
%%   <key>={<id1>,<id2>}                    %%
%%   <key>=                                 %%
%% Comment or delete the keys which are     %%
%% not used. Repeat \author command as much %%
%% as required.                             %%
%%                                          %%
%%%%%%%%%%%%%%%%%%%%%%%%%%%%%%%%%%%%%%%%%%%%%%

\author[
   addressref={aff1,aff2},                   % id's of addresses, e.g. {aff1,aff2}
   noteref={n1}                        % id's of article notes, if any
]{\inits{YM}\fnm{Yoichi} \snm{Miyahara}}
\author[
   addressref={aff1},
   noteref={n1}
]{\inits{HG}\fnm{Harrisonn} \snm{Griffin}}
\author[
   addressref={aff1}
]{\inits{ARG}\fnm{Antoine} \snm{Roy-Gobeil}}
\author[
   addressref={aff1}
]{\inits{ARG}\fnm{Ron} \snm{Belyansky}}
\author[
   addressref={aff1}
]{\inits{ARG}\fnm{Hadallia} \snm{Bergeron}}
\author[
   addressref={aff1,aff3}
]{\inits{JB}\fnm{José} \snm{Bustamante}}
\author[
addressref={aff1}
]{\inits{PG}\fnm{Peter} \snm{Grutter}}

%%%%%%%%%%%%%%%%%%%%%%%%%%%%%%%%%%%%%%%%%%%%%%
%%                                          %%
%% Enter the authors' addresses here        %%
%%                                          %%
%% Repeat \address commands as much as      %%
%% required.                                %%
%%                                          %%
%%%%%%%%%%%%%%%%%%%%%%%%%%%%%%%%%%%%%%%%%%%%%%

\address[id=aff1]{%                           % unique id
  \orgname{Department of Physics, McGill University}, % university, etc
  \street{3600 rue University},                     %
  \postcode{H3A 2T8}                                % post or zip code
  \city{Montreal},                              % city
  \cny{Canada}                                    % country
}
\address[id=aff2]{%
  \orgname{Department of Physics, Texas State University},
  \street{601 University Drive},
  \postcode{78666}
  \city{San Marcos},
  \cny{USA}
}
\address[id=aff3]{%
  \orgname{Departamento de Fisica, Universidad San Francisco de Quito},
  \street{Diego de Robles S/N},
  \postcode{170901}
  \city{Quito},
  \cny{Ecuador}
}

%%%%%%%%%%%%%%%%%%%%%%%%%%%%%%%%%%%%%%%%%%%%%%
%%                                          %%
%% Enter short notes here                   %%
%%                                          %%
%% Short notes will be after addresses      %%
%% on first page.                           %%
%%                                          %%
%%%%%%%%%%%%%%%%%%%%%%%%%%%%%%%%%%%%%%%%%%%%%%

\begin{artnotes}
%\note{Sample of title note}     % note to the article
  \note[id=n1]{Equal contributor} % note, connected to author
%  \note[id=n2]{email:yoichi.miyahara@txstate.edu} % note, connected to author
%  \note[id=n3]{email:peter.grutter@mcgill.ca} % note, connected to author
\end{artnotes}

%\end{fmbox}% comment this for two column layout

%%%%%%%%%%%%%%%%%%%%%%%%%%%%%%%%%%%%%%%%%%%%%%
%%                                          %%
%% The Abstract begins here                 %%
%%                                          %%
%% Please refer to the Instructions for     %%
%% authors on http://www.biomedcentral.com  %%
%% and include the section headings         %%
%% accordingly for your article type.       %%
%%                                          %%
%%%%%%%%%%%%%%%%%%%%%%%%%%%%%%%%%%%%%%%%%%%%%%

\begin{abstractbox}
\begin{abstract} % abstract
  Reliable operation of frequency modulation mode atomic force microscopy
  (FM-AFM) depends on a clean resonance of an AFM cantilever.
  It is recognized that the spurious mechanical resonances which originate from various mechanical components in the microscope body
  are excited by a piezoelectric element
  that is intended for exciting the AFM cantilever oscillation
  and these spurious resonance modes cause the serious undesirable signal artifacts in both frequency shift and dissipation signals.
  We present an experimental setup to excite only
  the oscillation of the AFM cantilever in a fiber-optic interferometer system using optical excitation force.
  While the optical excitation force is provided by a separate laser light source with a different wavelength (excitation laser : $\lambda = 1310$~nm),
  the excitation laser light is still guided through the same single-mode optical fiber
  that guides the laser light (detection laser : $\lambda = 1550$~nm) used
  for the interferometric detection of the cantilever deflection.
  We present the details of the instrumentation and
  its performance.
  This setup allows us to eliminate the problems associated
  with the spurious mechanical resonances
  such as the apparent dissipation signal
  and the inaccuracy in the resonance frequency measurement.
\end{abstract}

%%%%%%%%%%%%%%%%%%%%%%%%%%%%%%%%%%%%%%%%%%%%%%
%%                                          %%
%% The keywords begin here                  %%
%%                                          %%
%% Put each keyword in separate \kwd{}.     %%
%%                                          %%
%%%%%%%%%%%%%%%%%%%%%%%%%%%%%%%%%%%%%%%%%%%%%%


% MSC classifications codes, if any
%\begin{keyword}[class=AMS]
%\kwd[Primary ]{}
%\kwd{}
%\kwd[; secondary ]{}
%\end{keyword}

\end{abstractbox}
\end{fmbox}% uncomment this for twcolumn layout

\end{frontmatter}

%%%%%%%%%%%%%%%%%%%%%%%%%%%%%%%%%%%%%%%%%%%%%%
%%                                          %%
%% The Main Body begins here                %%
%%                                          %%
%% Please refer to the instructions for     %%
%% authors on:                              %%
%% http://www.biomedcentral.com/info/authors%%
%% and include the section headings         %%
%% accordingly for your article type.       %%
%%                                          %%
%% See the Results and Discussion section   %%
%% for details on how to create sub-sections%%
%%                                          %%
%% use \cite{...} to cite references        %%
%%  \cite{koon} and                         %%
%%  \cite{oreg,khar,zvai,xjon,schn,pond}    %%
%%  \nocite{smith,marg,hunn,advi,koha,mouse}%%
%%                                          %%
%%%%%%%%%%%%%%%%%%%%%%%%%%%%%%%%%%%%%%%%%%%%%%

%%%%%%%%%%%%%%%%%%%%%%%%% start of article main body
% <put your article body there>

%%%%%%%%%%%%%%%%
%% Background %%
%%
% \section*{Content}
% Text and results for this section, as per the individual journal's instructions for authors. %\cite{koon,oreg,khar,zvai,xjon,schn,pond,smith,marg,hunn,advi,koha,mouse}

\section*{Introduction}
Reliable operation of frequency modulation mode atomic force microscopy
(FM-AFM) depends on a clean resonance of an AFM cantilever.
It is recognized that
the spurious mechanical resonances originating from
the various mechanical components of the microscope body
are excited inadvertently by a piezoelectric element
that is intended for exciting the AFM cantilever oscillation
and these spurious resonances cause the serious undesirable signal artifacts
in both frequency shift and dissipation signals \cite{Kobayashi2011,Labuda2011}.

The additional frequency-dependent amplitude response causes
a crosstalk between the dissipation and frequency shift signals,
resulting in artifact signal in dissipation channel
which misleads the interpretation of the tip-sample interaction physics.
Although such dissipation artifacts could be corrected for
by off-line signal processing \cite{Labuda2011},
the additional frequency-dependent phase responses
near the cantilever's resonance frequency
can cause the deviation of the oscillation frequency
from the cantilever's resonance frequency
for the cantilevers with low quality (Q) factor case \cite{Kobayashi2011}
which makes quantitative measurement of tip-sample interaction challenging.
Even in the case of the high-Q cantilever,
excessive amount of phase changes due to spurious resonances can cause
an interruption of the self oscillation by breaking the phase matching condition,
resulting in the loss of distance feedback signal.
For this reason, it is important to develop a cantilever excitation scheme
which can eliminate the excitation of the spurious mechanical resonances.
While the optical excitation of AFM cantilever oscillation has been adopted mainly
for the operation in liquid in which case the quality factor
is strongly damped \cite{Labuda2012a},
it is also widely recognized 
that the effects of spurious resonances are important at low temperature in high vacuum environment \cite{Labuda2011}
where the Q-factors of spurious resonances are enhanced substantially.
Here we report an experimental setup for optical AFM cantilever excitation
based on an all fiber optic interferometer which is commonly used for
low-temperature AFM systems \cite{Rugar1989}.
As this setup requires only one optical fiber which is used for both
sensing and actuation, it can be easily adopted to the existing AFMs
which employ fiber-optic interferometers.  

\section*{Experimental}
\begin{figure}[h]
  \centering
  \includegraphics[width=10cm]{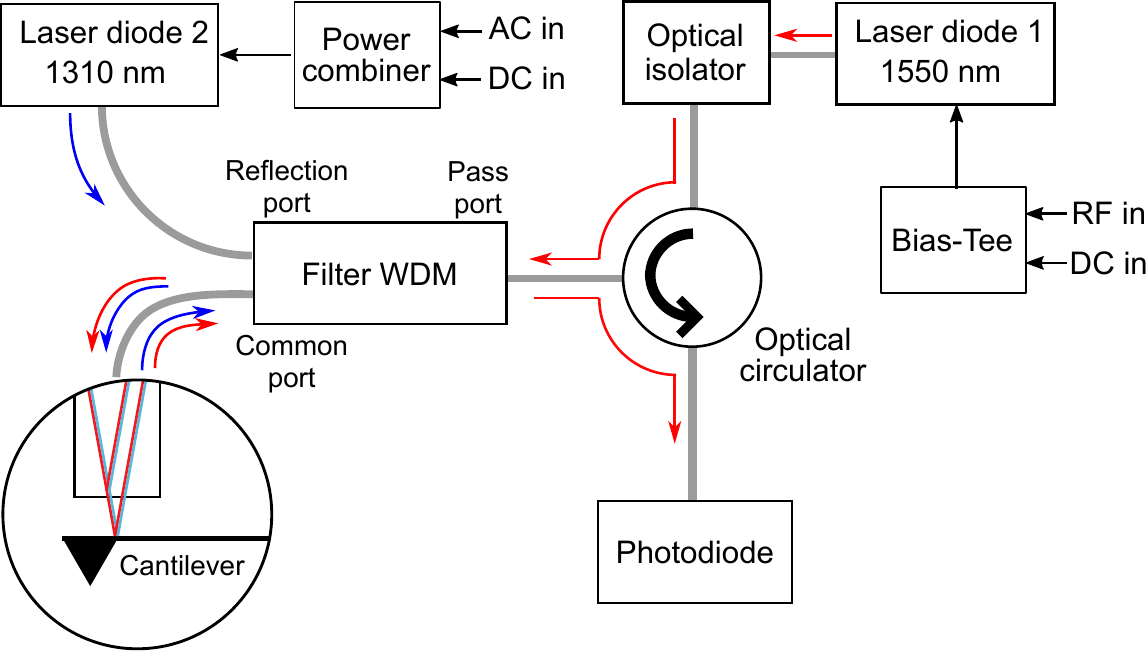}\\
  \caption{Schematic diagram of the experimental setup.
    Two laser lights from Laser diode 1 (1550~nm) and Laser diode 2 (1310~nm)
    are combined with a filter Wavelength Division Multiplexer (FWDM).
    The combined light is launched into a single-mode optical fiber.
    While a fraction of the light is reflected at the fiber-vacuum interface,
    the rest of the light which  comes out of the fiber is reflected
    on the back side of the cantilever and goes back into the fiber.
    Only the 1550~nm laser can pass through the filter WDM and reach
    the photo diode.
    \label{setup}}
\end{figure}

Figure~\ref{setup} shows a diagram of the optical setup.
In this setup, two separate laser lights with different wavelengths
are used for the detection
(Laser diode 1, 1550~nm DFB laser with an optical isolater, NECSEL) and excitation (Laser diode 2, 1310~nm, LPS-1310-FC, Thorlabs)
of the AFM cantilever oscillation.
The excitation and detection laser lights are combined
with a filter wavelength division multiplexer
(FWDM, FWDM-1513, AFW Technology)
and launched into a single-mode optical fiber (SMF-28e) \cite{Weld2006}.
The reflected light from the fiber end and cantilever interfere each other
and go back the same FWDM.
Only the detection laser light can pass the FWDM and reach a photodiode
via an optical circulator (CIR-3-15, AFW Technology) \cite{Rasool2010}.
The intensity of the excitation laser is modulated by modulating the drive current with a power combiner (PRSC-2050, Mini Circuits).
The drive current for the detection laser is modulated by a radio frequency
signal (several hundred MHz) to suppress stray interferences \cite{Smith09}.

\begin{figure}[h]
  \centering
  \includegraphics[width=8.5cm]{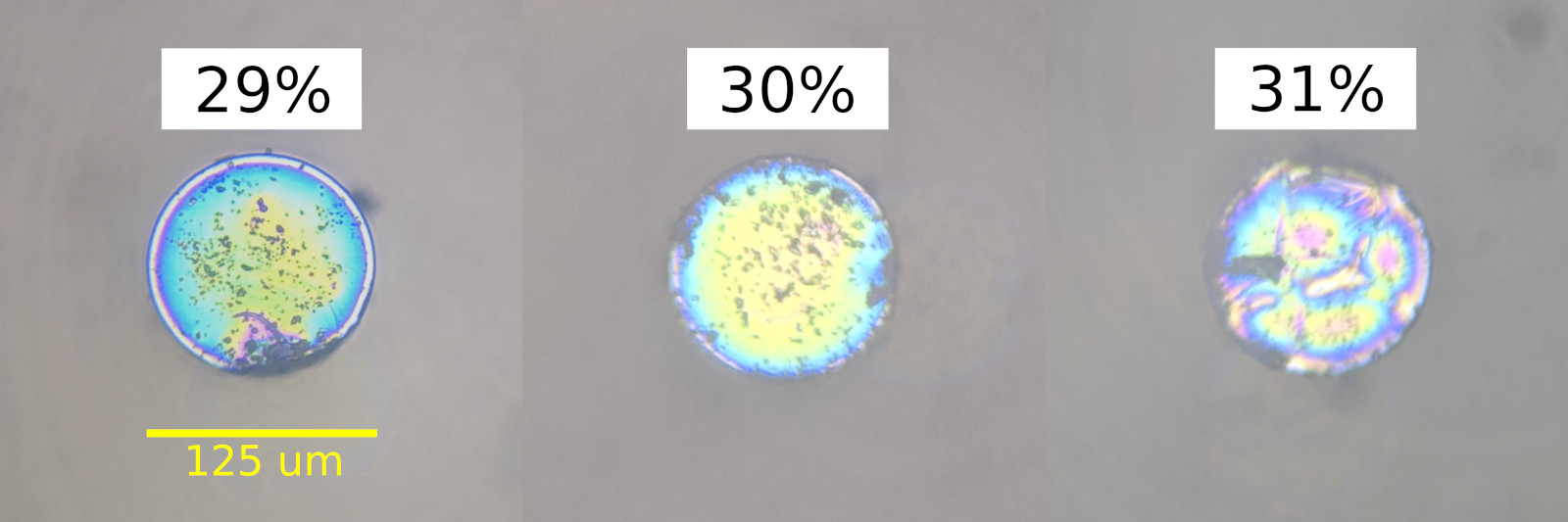}\\
  \caption{
    Three example fibers showing the reflective titanium dioxide coating with varying back reflectance.
    The fiber has a core diameter of $8~\mu$m with a fiber cladding of $125~\mu$m radius.
    \label{fiber_ends}}
\end{figure}

We coated the cleaved optical fiber end to increase the back reflection
to improve the detection sensitivity of the interferometer \cite{Subba-Rao09}.
The $\mathrm{TiO_2}$ solution used for fiber coating was prepared using a solution of titanium-(IV)-2-ethylhexoxide (Sigma-Aldrich) diluted with xylene.
The cleaved fiber end was dipped 1~cm into the solution, then passed through a propane torch for less than one second for flash annealing. Optimal flash parameters, such as placement of fiber in the flame and time in flame were determined via trial and error. The reflectivity of the fiber was monitored in real time with a photodiode back reflection set up. Successfully annealed fibers would typically have reflectivity between 20 and 30\%, with the best fibers having up to 35\%. Examples of successfully coated fibers are shown in Fig~\ref{fiber_ends}.

%The back reflection at the fiber end was determined by comparing the measured reference and reflected beam via two photodiode detectors. Only bare fibers that had back reflection of at least 3\% were used for the thin film coating.

\section*{Characterizing optical cavity}
The home built low temperature AFM uses a Fabry-Perot interferometer (FPI)
in reflective mode to precisely and accurately 
measure the distance between the end of the optical fiber
and the reflective aluminum coating of the AFM cantilever.
An FPI is an optical cavity which is made with two parallel semi-reflective mirrors
(mirror 1 and 2) with reflectances $R_1$ and $R_2$, separated by a cavity length $d$.
In our case, the coated fiber and cantilever are the mirror $1$ and $2$, respectively,
and the cavity medium is vacuum.
The coated fiber is aligned to the tip end of the cantilever 
to maximize the deflection measurement sensitivity.
Incident light from the detection laser travels down the fiber,
where some of the light is internally reflected off of the fiber end,
and the rest is injected into the FPI.
This light is reflected off of the cantilever
where it is then reflected multiple times between the two mirrors of the cavity.
Each time the light beam encounters the coated fiber end,
it is partially reflected back into the cavity
while the rest enters back into the fiber.
All the transmitted beams into the fiber together with the initially reflected beam
contribute to the interference inside the fiber.
The reentered light is guided through the fiber, WDM, and circulator
and its intensity, $I_r$, is measured by the photodiode (Fig.~\ref{setup}).
The FPI lends itself to being more sensitive than other interferometers
because of the multiple reflections between the mirror 1 and 2 
when the separation between the mirrors is small enough \cite{Vogel2003, Wilkinson2011, VonSchmidsfeld2015}.
It is important to notice that the multiple reflection
also plays an important role for the excitation laser light 
as it enhances the optical force.
We use the FPI model reported in Ref.~\cite{Ismail2016} 
for the following analysis.

The ratio of the intracavity intensity, $I_\mathrm{circ}$,
to the intensity incident upon mirror 1 , $I_\mathrm{inc}$,
(enhancement factor, $\Acirc'$) is given by
Airy distribution,

\begin{equation}
  \Acirc ' = \frac{\Icirc}{\Iinc} = \frac{1-R_1}{(1-\sqrt{R_1 R_2})^2 + 4\sqrt{R_1 R_2} \sin^2 \phi}.
\end{equation}
where $\phi$ is the single-pass phase shift between the mirrors
and can be expressed as
%resulting phase delay from the multiple reflections of $I_r$ inside of the cavity. 
%We can express $\phi$ as
\begin{equation}
  \label{eq:phi}
  \phi = \frac{2 \pi d}{\lambda} %\left( z \cos{\theta_t + 2 \delta} \right)
\end{equation}
where $d$ is the distance between two mirrors (cavity length) and $\lambda$ is wavelength.
Similarly, the enhancement factor for
the total back reflected light with respect to $I_\mathrm{inc}$
is given by

\begin{equation}
  \label{eq:reflectance}
  \frac{\Irefl}{\Iinc} = \frac{(\sqrt{R_1}-\sqrt{R_2})^2 + 4\sqrt{R_1 R_2} \sin^2\phi}
  {(1-\sqrt{R_1 R_2})^2 + 4\sqrt{R_1 R_2} \sin^2 \phi}.
  %  = \frac{F (\alpha  +  \sin^2{\phi})}
  %  {1 + F \sin^2{\phi}}
\end{equation}

\begin{figure}[h]
  \centering
  \includegraphics[width=8.5cm]{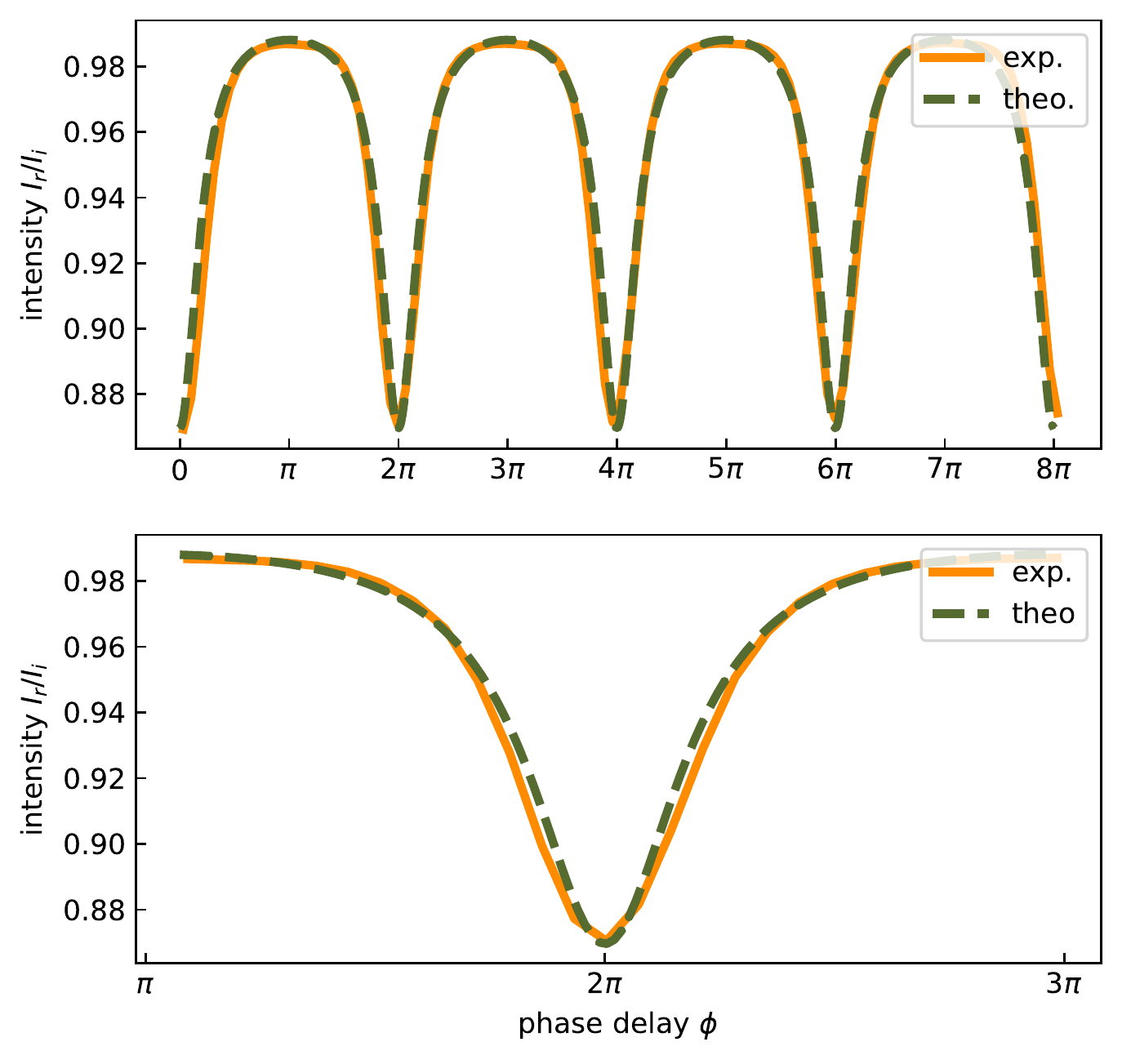}\\
  \caption{ Theoretical and experimental transfer function of the reflective mode
    of the Fabry-Perot interferometer.
    \textbf{Top} shows the measured and theoretical interference response
    of the cavity as a function of increasing phase delay
    (i.e. increasing fiber-cantilever separation).
    This fiber end has a 30\% back reflection.
    \textbf{Bottom} shows the first full interference peak that occurs
    as peak intensity as a function of phase delay.
    The experimental data is plotted with the theoretical fit.
    \label{cavity_model_phase}}
\end{figure}

Figure~\ref{cavity_model_phase}(a) shows the experimentally measured interference fringe and
theoretical one given by Eq.~\ref{eq:reflectance} assuming $R_1 = 0.33$ and $R_2=0.96$.
These values are in good agreement with the value obtained
from the measured back reflection of the coated fiber ($R_1 = 0.30$)
and the reflectance of the Al coated cantilever $R_2 = 0.96$
assuming 100\% collection of the reflected beam from the cantilever.
The good agreement between the theory and experiment confirms
the validity of the model.
For $R_1=0.33$ and $R_2=0.96$, the enhancement factor, $\Acirc'$ is found to
be 3.5.

\section*{Results}
\begin{figure}[h]
  \centering
  \includegraphics[width=10cm]{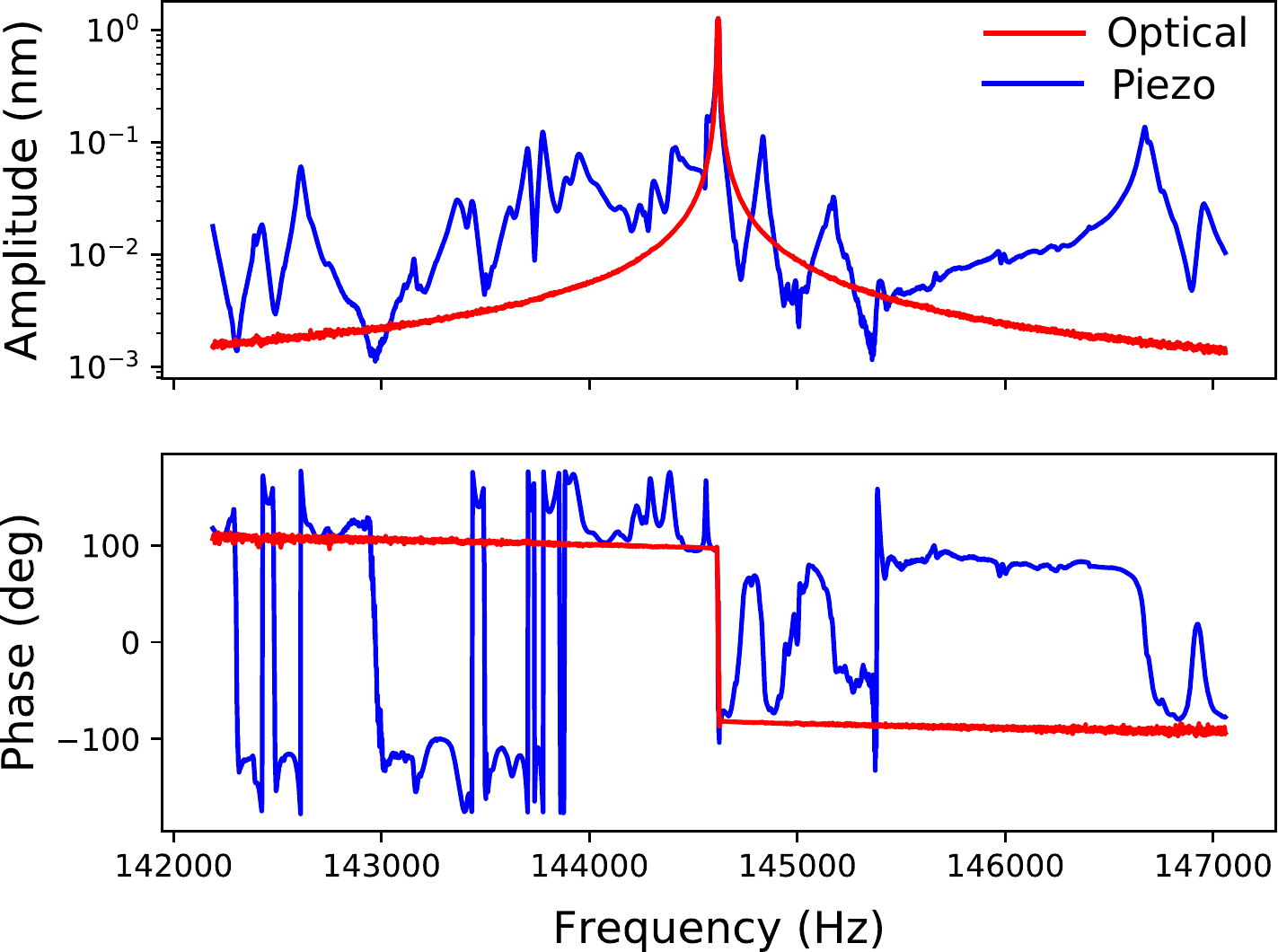}\\
  \caption{ Amplitude and phase frequency responses of the AFM cantilever excited
    by optical excitation (red) and piezo excitation (blue) measured at 4.5 K.
    Note the spurious mechanical resonances in the piezo excitation which arise from vibrations in the microscope body.
    \label{comparison_spectra}}
\end{figure}

Figure~\ref{comparison_spectra} shows
the frequency responses of an AFM cantilever
(NCLR Nanosensors, Al backside coated and Pt tip-side coated)
excited by piezo excitation (blue line) and optical excitation (red line).
The response by piezo excitation shows spurious resonance modes
over the entire 5~kHz frequency range around the cantilever's fundamental resonance frequency.
These additional frequency-dependent amplitude and phase responses are not present
when the cantilever is optically driven.
The optical force does not excite the various mechanical resonances
of the microscope body,
thus giving us cleaner signals in the frequency shift and dissipation channels.
Through removing the artifacts, we can better study the tip-sample interaction physics.

\subsection*{Quantifying optical excitation force}
In order to determine the magnitude of the optical force quantitatively,
we use the following relation between the oscillation amplitude
and the amplitude of the excitation optical force
when the cantilever is excited by harmonic excitation force,
$F(t)=\Fopt \sin(2\pi f_0 t)$,
at the resonance frequency, $f_0$:
\begin{equation}
  \label{eq:amplitude}
  A = \frac{Q}{k} \Fopt.
\end{equation}
Here $A$ is the amplitude of the oscillation,
$Q$ the quality factor, $k$ the spring constant of the cantilever
and $\Fopt$ the amplitude of the oscillating optical force,
We used the Sader method to calibrate the effective spring constant
of the fundamental flexural mode of cantilevers \cite{sader1999calibration}
and obtained a spring constant $k=19$~N/m for the cantilever used.

In our interferometer setup,
the fiber-cantilever distance can be adjusted by a piezo-electric stick-slip motor
with the step size as small as $20 ~ \mathrm{nm}$.
The step size is confirmed to be very uniform over several fringes.
As we change the fiber-cantilever distance by stepping the fiber position,
we record the signal at the photodiode $\Vpd$.
The resulting interference fringes show nearly periodic peaks
and the peak separation, known to be $\lambda / 2$
from Eq.~\ref{eq:reflectance}.
This well-defined separation allows us to calibrate the horizontal axis 
that determine the sensitivity of the interferometer.

\begin{figure}[h]
  \centering
  \includegraphics[width=10cm]{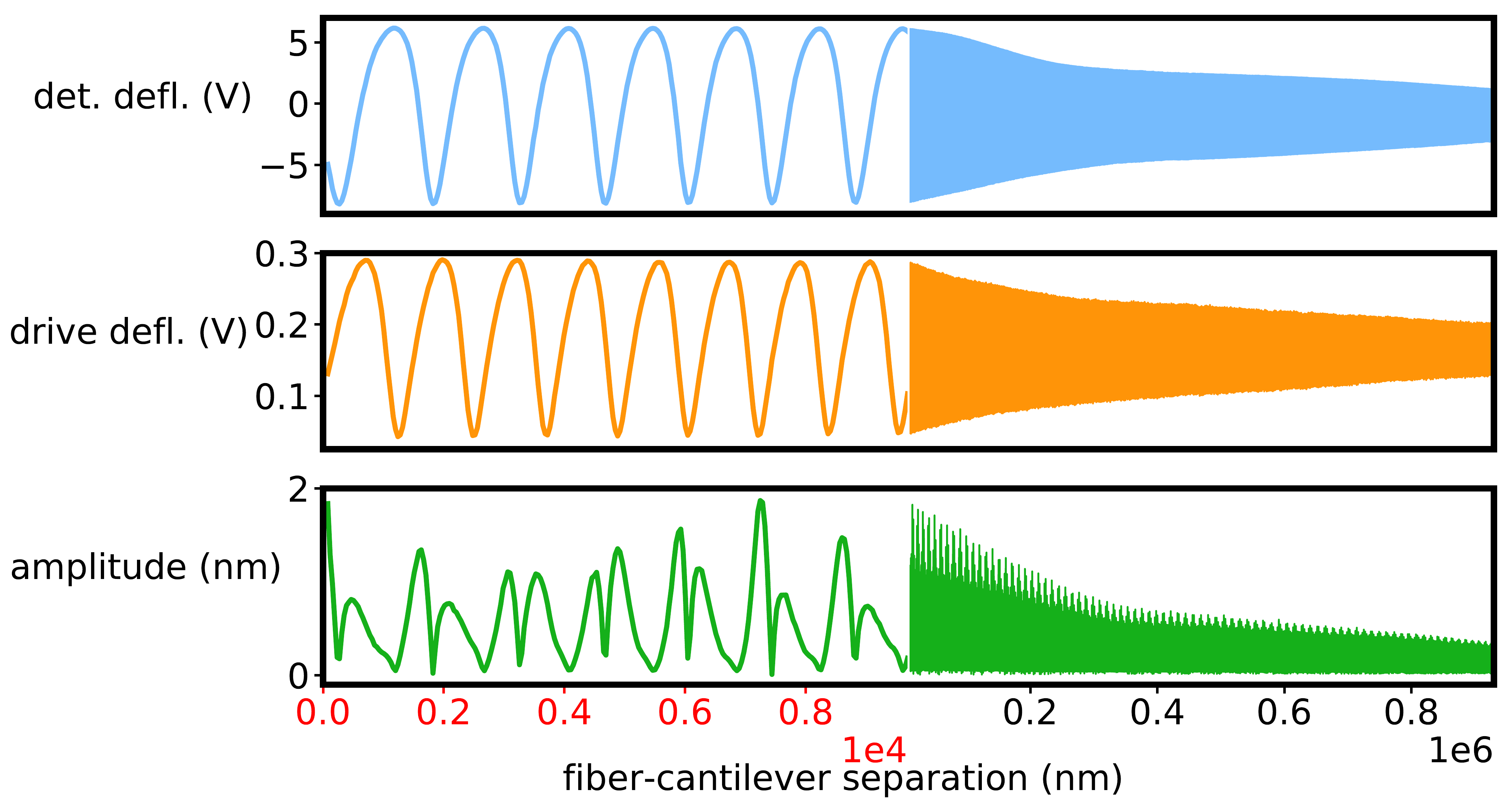}\\
  \caption{Interference fringes of detection (1550~nm) (cyan) and excitation (1310~nm) (orange) lasers
    as a function of fiber-cantilever separation measured at 77 K in vacuum of $10^{-5}$ mbar.
    \label{cavity_scan}}
\end{figure}

\begin{figure}[h]
  \centering
  \includegraphics[width=10cm]{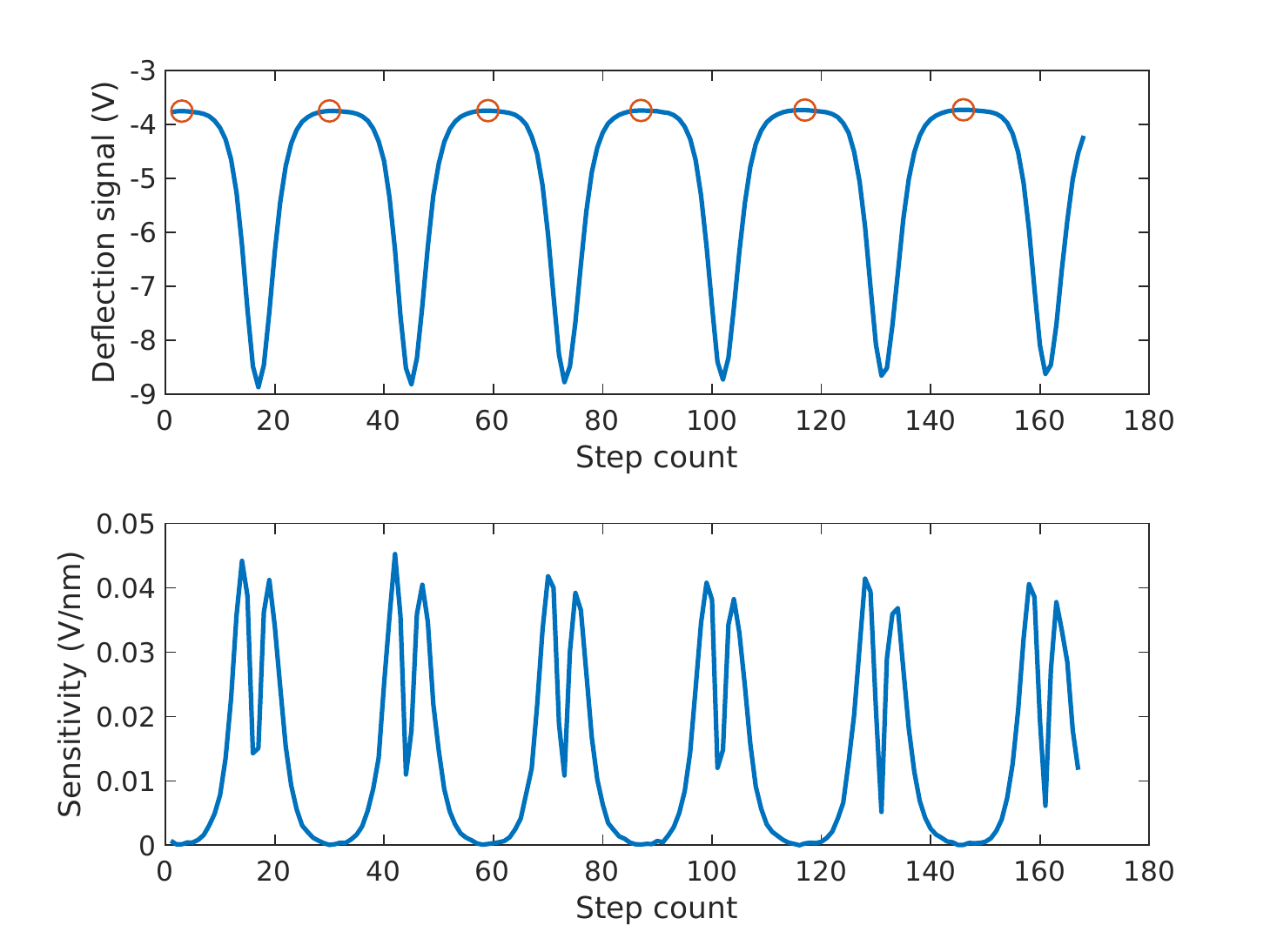}\\
  \caption{
    \textbf{Top} shows the photodiode signal as a function of the number of fiber walker steps
    as it is retracted from the cantilever.
    Peaks are indicated via circles, which are then used to determine the number of steps between adjacent fringes.
    \textbf{Bottom} shows sensitivity $S$ which is calculated
    by taking the derivative of the photodiode signal and dividing by the average step size between fringes.
    \label{sensitivity}}
\end{figure}

The interferometer sensitivity $S$ is calculated
as the derivative of the measured interferometer output voltage $\Vpd$
with respect to the fiber fiber-cantilever distance $z$ \cite{antrg2017}:
Since the distance between successive peaks is known,
we can convert the walker steps $\Delta z$ into a distance
by identifying successive peak positions
and counting the steps between them.
Now we can simply take the derivative of the photodiode signal versus the fiber-cantilever distance,
then divide by the walker step $\Delta z$ to obtain the sensitivity $S$.
An example of such measurement is shown in Fig.~\ref{sensitivity}.
With the known sensitivity in units of $\mathrm{V} / \mathrm{nm}$,
it is straightforward to determine the oscillation amplitude of the cantilever.
We measure the ac component of the photodiode signal $\Vpd$,
and convert its amplitude in Volt to the amplitude in nm
by dividing $\Vpd$ by the sensitivity $S$.
The typical sensitivity during the experiments is $30$~mV/nm.
This sensitivity can also be used to calculate the detection noise
of the interferometer.
By taking a power spectral density of $\Vpd$ around the cantilever resonance,
we can convert the measured interferometer noise in $\mathrm{V}/\sqrt{\mathrm{Hz}}$
into the detection noise in $\mathrm{fm}/ \mathrm{\sqrt{Hz}}$.
A typical detection noise at 77~K is $\approx 50~ \mathrm{fm} / \sqrt{\mathrm{Hz}}$
and at 4~K is $ \approx 15 ~\mathrm{fm} / \sqrt{\mathrm{Hz}}$.
The best detection noise we have observed at 4~K is $7~\mathrm{fm}/\mathrm{\sqrt{Hz}}$.

\begin{figure}[h]
  \centering
  \includegraphics[width=8.2cm]{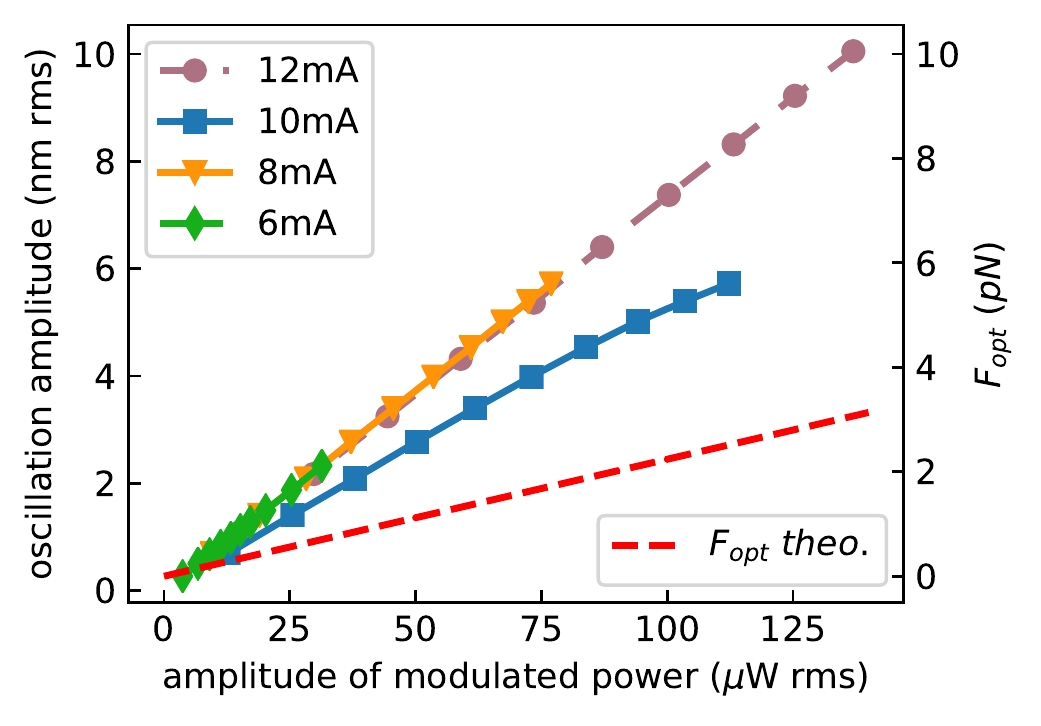}\\
  \caption{ Cantilever oscillation amplitude
    as a function of modulation power at varying laser diode currents
    (solid lines).
    The right vertical axis indicates the corresponding optical force.
    The red dashed line shows the theoretical optical scattering force.
    \label{modulation_power}
  }
\end{figure}

From the measured $A$, $Q$ and $k$,
we are able to obtain the optical excitation force, $\Fopt$
as a function of the amplitude of the modulated optical power by using Eq.~\ref{eq:amplitude},
as shown on the right hand side axis of Fig.~\ref{modulation_power}.
We measured the optical power of the $1310$~nm driving laser
with a separate photodiode placed at the 10~\% branch of a 90/10 coupler
(not shown in Fig.~\ref{setup})
and its modulated amplitude, $P$, with the lock-in amplifier.
Figure~\ref{modulation_power} shows the measured $A$-$P$ relations
with four different dc offset current settings.
We can see that the oscillation amplitude is just dependent on
the amplitude of modulated optical power and does not depend
on the dc power.
The average slope of the four data sets is $68.4$~pN/mW.

Now let us compare the measured driving force with the theoretical optical forces.
The theoretical optical scattering force (radiation pressure force), $\Fscat$,
acting on the AFM cantilever with reflectance $R_2$ in a FPI system is given by
\begin{equation}
  \label{eq:scattering_force}
  \Fscat = \frac{2 \Pcirc R_2}{c} = \frac{2 \Acirc' \Pinc R_2}{c}
\end{equation}
where $\Pcirc$ is the optical power in the FPI cavity,
$\Pinc$ the incident optical power and
$c$ the speed of light.
While the enhancement factor $\Acirc'$ is in general dependent on
the wavelength due to the wavelength-dependence of $R_2$,
we confirmed $\Acirc'=3.5$ for the excitation laser ($\lambda = 1330$~nm)
by measuring the interferometer fringes as shown in Fig.~\ref{cavity_scan}.
The theoretical slope of $\Fscat$-$\Pinc$ is therefore
$\Fscat/ \Pinc=22.4$~pN/mW for $\Acirc'=3.5$ and $R_2 = 0.96$.
The fact that the experimental optical force $\Fopt$ is much larger
than the expected optical scattering force $\Fscat$
indicates that the photothermal (bolometric) effect plays an important role
\cite{Vogel2003}.

\begin{figure}[h]
  \centering
  \includegraphics[width=8.2cm]{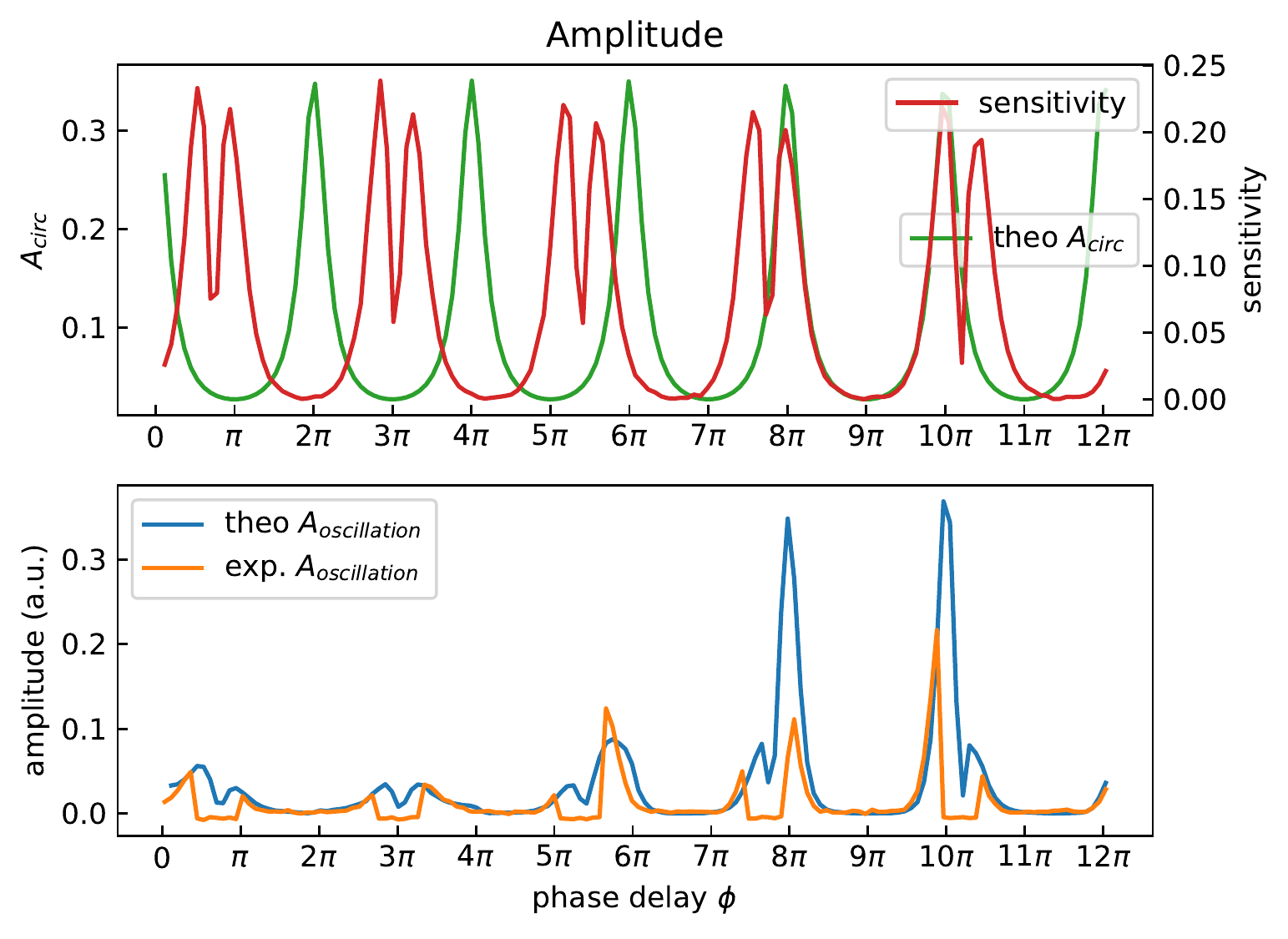}\\
  \caption{ \textbf{Top} shows the theoretical $\Acirc'$
    and the sensitivity $S$ for a set of interference fringes.
    \textbf{Bottom} shows the theoretical and experimentally measured oscillation amplitude
    of the cantilever as a function of phase delay (cavity length).
    The theoretical amplitude was calculated
    by multiplying circulating enhancement factor $\Acirc'$ by sensitivity $S$.
    \label{Comparison_amplitude}}
\end{figure}

\subsection*{Effect of optical driving force on frequency shift noise}
In FM-AFM, the interaction forces between the sample and cantilever 
are detected as a frequency shift. 
Since the tip-sample distance is regulated by the frequency shift, 
a reduction in frequency shift noise will lead to high-resolution
FM-AFM images. 
Additionally, in electrostatic force microscopy
with single-electron sensitivity (e-EFM) measurements,
achieving a detectable signal is only possible with a
reduced frequency shift noise \cite{antrg2017, Roy-Gobeil2019}. 
In the following we will investigate
if optically driving the cantilever introduces additional frequency noise. 
The frequency shift noise measurements were performed at 77~K
with the AFM cantilevers that have spring constants of $k = 20$~N/m
and quality factors at 77~K varying between 15,000 and 50,000.
In these experiments, the cantilever is self-excited at its resonant frequency
by feeding the deflection signal into the optical excitation system
through an oscillation control electronics (easyPLLplus controller, Nanosurf)
which consists of a phase shifter and amplitude controller \cite{Albrecht1991}.

\begin{figure}[h]
  \centering
  \includegraphics[width=8.2cm]{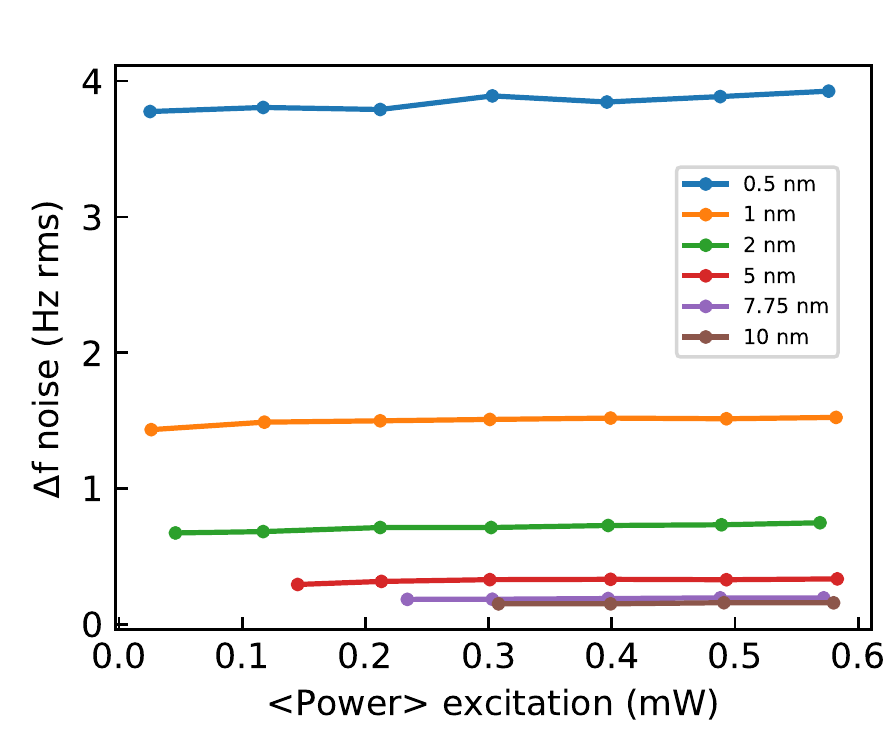}\\
  \caption{ Frequency shift noise as a function of average optical power
    of the excitation laser at varying oscillation amplitudes.
    The frequency shift noise was measured as the root mean square (RMS)
    of the frequency shift from the lock-in amplifier.
    The lines are a guide to the eyes.
    \label{df_v_power}
  }
\end{figure}

To see if the average power of the excitation source
affected the frequency shift noise of the microscope,
we measured the frequency shift noise at various oscillation amplitudes 
from 0.5 to 10~nm,
and increased the average optical power of the excitation laser.
The amplitude was kept constant at different average optical powers. 
The frequency shift measurement was made using a phase-locked loop (PLL)
frequency detector (HF2LI with PLL option, Zurich Instruments) 
with a detection bandwidth of 100~Hz.
The frequency shift noise was measured using the root mean square value of the frequency shift
and is shown in Fig.~\ref{df_v_power}.
As expected from the theory \cite{Albrecht1991},
the frequency shift noise increases with decreasing oscillation amplitude,
but there is no significant variation in the frequency shift noise
at the same oscillation amplitude for varying average optical excitation power.

\begin{figure}[h]
  \centering
  \includegraphics[width=7cm]{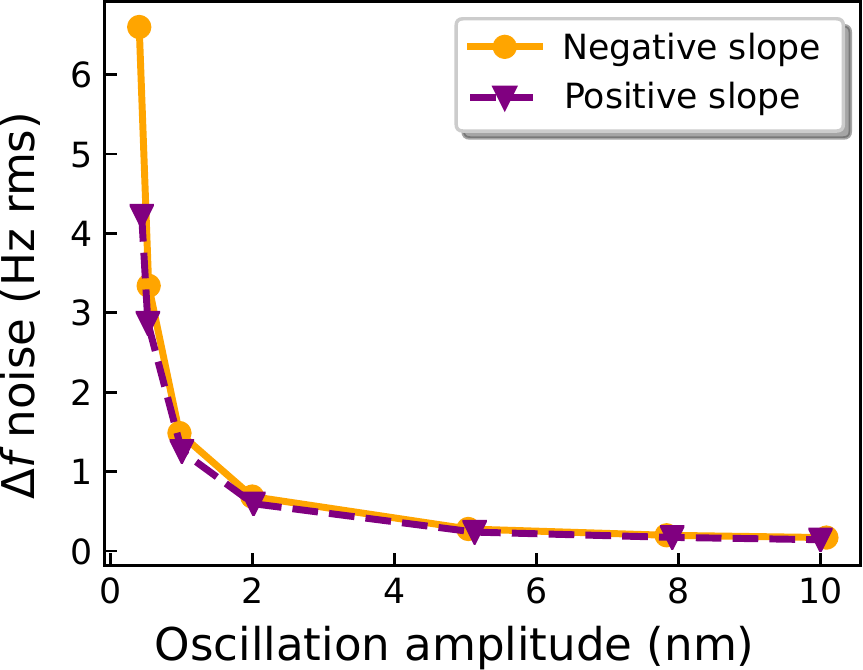}\\
  \caption{ Frequency shift noise ($\Delta f$) as a function of oscillation amplitude of cantilever
    being driven via optical excitation at 77 K.
    $\Delta f$ is measured on the negative slope (orange circle) and positive slope (purple triangle)
    of the interference fringe.
    The lines are a guide to the eyes.
    \label{df_v_osc}
  }
\end{figure}

Since we can precisely adjust our fiber-cantilever separation distance,
we are able to choose an operating position on an interference fringe arbitrarily.
Typically, we choose a fringe position that maximizes the sensitivity.
On a single fringe, we could choose the operating position
either on the negative or positive slope side of the fringe.
It has been reported that the effective Q-factor is decreased or increased
depending on the slope of the interferometer signal
due to the optomechanical coupling \cite{Vogel2003, Holscher2009, Troger2010}.
For example, by adjusting the fiber-cantilever cavity position
to be on either side of a single fringe at the same deflection DC value,
we observe the effective Q-factors at the negative and positive slope side to vary as much as 3000.
To determine if the different effective Q-factor at the negative and positive slope
affect the frequency shift noise,
we measured the frequency shift noise at various oscillation amplitudes
on both the negative and positive slope side of the fringe at the same DC deflection values,
as shown in Fig.~\ref{df_v_osc}.
The results shows negligible difference in frequency shift noise
between the negative slope and positive slope side of the fringe.

\section*{Conclusion}
In conclusion, we have integrated an optical drive cantilever excitation system
into our low temperature AFM in a non-invasive way.
By using the optical circuit as shown in Fig.~\ref{setup}, 
we are able to have a two laser detection and excitation system 
deployed through a single optical fiber. 
The optical drive reduces spurious mechanical resonances 
making the AFM suitable to study sensitive tip-sample physics 
without artifacts from crosstalk in the frequency shift and dissipation channels.
The setup can easily be adopted for the existing AFMs 
which use fiber-optic interferometer.

\begin{backmatter}

\section*{Competing interests}
  The authors declare that they have no competing interests.

\section*{Author's contributions}
YM designed and constructed the experimental setup.
YM, HG, ARG and JB performed the experiments.
YM, HG and ARG analyzed the data.
RB and HB contributed to the preliminary experiments.
YM and PG conceived of the study.
YM, HG and PG wrote the manuscript
All authors read and approved the final manuscript.

\section*{Acknowledgements}
Financial support from The Natural Sciences and Engineering Research Council of Canada
and Le Fonds de Recherche du Québec—Nature et Technologies are gratefully acknowledged.
JB acknowledges financial support from National Secretary of Higher Education, Science and
Technology and Innovation, Ecuador.

\section*{Additional information}
Correspondence and requests for materials should be addressed to YM (yoichi.miyahara@txstate.edu)
or PG (peter.grutter@mcgill.ca).

%%  After submission of the .TEX file,                     %%
%%  you will be prompted to submit your .BBL file.         %%

% if your bibliography is in bibtex format, use those commands:
\bibliographystyle{bmc-mathphys} % Style BST file (bmc-mathphys, vancouver, spbasic).

\newcommand{\BMCxmlcomment}[1]{}

\BMCxmlcomment{

<refgrp>

<bibl id="B1">
  <title><p>{Reduction of frequency noise and frequency shift by phase shifting
  elements in frequency modulation atomic force microscopy}</p></title>
  <aug>
    <au><snm>Kobayashi</snm><fnm>K</fnm></au>
    <au><snm>Yamada</snm><fnm>H</fnm></au>
    <au><snm>Matsushige</snm><fnm>K</fnm></au>
  </aug>
  <source>Review of Scientific Instruments</source>
  <pubdate>2011</pubdate>
  <volume>82</volume>
  <issue>3</issue>
  <fpage>033702</fpage>
  <url>http://link.aip.org/link/RSINAK/v82/i3/p033702/s1{\&}Agg=doi
  http://scitation.aip.org/content/aip/journal/rsi/82/3/10.1063/1.3557416</url>
</bibl>

<bibl id="B2">
  <title><p>{Decoupling conservative and dissipative forces in frequency
  modulation atomic force microscopy}</p></title>
  <aug>
    <au><snm>Labuda</snm><fnm>A</fnm></au>
    <au><snm>Miyahara</snm><fnm>Y</fnm></au>
    <au><snm>Cockins</snm><fnm>L</fnm></au>
    <au><snm>Grutter</snm><fnm>PH</fnm></au>
  </aug>
  <source>Physical Review B</source>
  <publisher>American Physical Society</publisher>
  <pubdate>2011</pubdate>
  <volume>84</volume>
  <issue>12</issue>
  <fpage>125433</fpage>
  <url>http://link.aps.org/doi/10.1103/PhysRevB.84.125433
  https://link.aps.org/doi/10.1103/PhysRevB.84.125433</url>
</bibl>

<bibl id="B3">
  <title><p>{Retrofitting an atomic force microscope with photothermal
  excitation for a clean cantilever response in low Q environments}</p></title>
  <aug>
    <au><snm>Labuda</snm><fnm>A</fnm></au>
    <au><snm>Kobayashi</snm><fnm>K</fnm></au>
    <au><snm>Miyahara</snm><fnm>Y</fnm></au>
    <au><snm>Gr{\"{u}}tter</snm><fnm>P</fnm></au>
  </aug>
  <source>Review of Scientific Instruments</source>
  <pubdate>2012</pubdate>
  <volume>83</volume>
  <issue>May</issue>
  <fpage>053703</fpage>
  <url>http://dx.doi.org/10.1063/1.4712286</url>
</bibl>

<bibl id="B4">
  <title><p>{Improved fiber‐optic interferometer for atomic force
  microscopy}</p></title>
  <aug>
    <au><snm>Rugar</snm><fnm>D.</fnm></au>
    <au><snm>Mamin</snm><fnm>H. J.</fnm></au>
    <au><snm>Guethner</snm><fnm>P.</fnm></au>
  </aug>
  <source>Applied Physics Letters</source>
  <pubdate>1989</pubdate>
  <volume>55</volume>
  <issue>25</issue>
  <fpage>2588</fpage>
  <lpage>-2590</lpage>
  <url>http://aip.scitation.org/doi/10.1063/1.101987</url>
</bibl>

<bibl id="B5">
  <title><p>{Feedback control and characterization of a microcantilever using
  optical radiation pressure}</p></title>
  <aug>
    <au><snm>Weld</snm><fnm>DM</fnm></au>
    <au><snm>Kapitulnik</snm><fnm>A</fnm></au>
  </aug>
  <source>Applied Physics Letters</source>
  <pubdate>2006</pubdate>
  <volume>89</volume>
  <issue>16</issue>
  <fpage>164102</fpage>
  <url>http://aip.scitation.org/doi/10.1063/1.2362598</url>
</bibl>

<bibl id="B6">
  <title><p>{A low noise all-fiber interferometer for high resolution frequency
  modulated atomic force microscopy imaging in liquids}</p></title>
  <aug>
    <au><snm>Rasool</snm><fnm>HI</fnm></au>
    <au><snm>Wilkinson</snm><fnm>PR</fnm></au>
    <au><snm>Stieg</snm><fnm>AZ</fnm></au>
    <au><snm>Gimzewski</snm><fnm>JK</fnm></au>
  </aug>
  <source>Review of Scientific Instruments</source>
  <pubdate>2010</pubdate>
  <volume>81</volume>
  <issue>2</issue>
  <fpage>023703</fpage>
  <url>http://www.ncbi.nlm.nih.gov/pubmed/20192498
  http://aip.scitation.org/doi/10.1063/1.3297901</url>
</bibl>

<bibl id="B7">
  <title><p>{A fiber-optic interferometer with subpicometer resolution for dc
  and low-frequency displacement measurement}</p></title>
  <aug>
    <au><snm>Smith</snm><fnm>D T</fnm></au>
    <au><snm>Pratt</snm><fnm>J R</fnm></au>
    <au><snm>Howard</snm><fnm>L P</fnm></au>
  </aug>
  <source>Review of Scientific Instruments</source>
  <publisher>AIP</publisher>
  <pubdate>2009</pubdate>
  <volume>80</volume>
  <issue>3</issue>
  <fpage>035105</fpage>
  <url>http://aip.scitation.org/doi/10.1063/1.3097187</url>
</bibl>

<bibl id="B8">
  <title><p>{Improving a high-resolution fiber-optic interferometer through
  deposition of a TiO2 reflective coating by simple dip-coating}</p></title>
  <aug>
    <au><snm>Subba Rao</snm><fnm>V</fnm></au>
    <au><snm>Sudakar</snm><fnm>C</fnm></au>
    <au><snm>Esmacher</snm><fnm>J</fnm></au>
    <au><snm>Pantea</snm><fnm>M</fnm></au>
    <au><snm>Naik</snm><fnm>R</fnm></au>
    <au><snm>Hoffmann</snm><fnm>PM</fnm></au>
  </aug>
  <source>Review of Scientific Instruments</source>
  <publisher>AIP</publisher>
  <pubdate>2009</pubdate>
  <volume>80</volume>
  <issue>11</issue>
  <fpage>115104</fpage>
  <url>http://link.aip.org/link/?RSI/80/115104/1
  http://aip.scitation.org/doi/10.1063/1.3244088</url>
</bibl>

<bibl id="B9">
  <title><p>{Optically tunable mechanics of microlevers}</p></title>
  <aug>
    <au><snm>Vogel</snm><fnm>M.</fnm></au>
    <au><snm>Mooser</snm><fnm>C.</fnm></au>
    <au><snm>Karrai</snm><fnm>K.</fnm></au>
    <au><snm>Warburton</snm><fnm>R. J.</fnm></au>
  </aug>
  <source>Applied Physics Letters</source>
  <pubdate>2003</pubdate>
  <volume>83</volume>
  <issue>7</issue>
  <fpage>1337</fpage>
  <lpage>-1339</lpage>
  <url>http://aip.scitation.org/doi/10.1063/1.1600513</url>
</bibl>

<bibl id="B10">
  <title><p>{Analytical model for low finesse, external cavity, fiber
  Fabry–Perot interferometers including multiple reflections and angular
  misalignment}</p></title>
  <aug>
    <au><snm>Wilkinson</snm><fnm>PR</fnm></au>
    <au><snm>Pratt</snm><fnm>JR</fnm></au>
  </aug>
  <source>Applied Optics</source>
  <pubdate>2011</pubdate>
  <volume>50</volume>
  <issue>23</issue>
  <fpage>4671</fpage>
  <url>https://www.osapublishing.org/abstract.cfm?URI=ao-50-23-4671</url>
</bibl>

<bibl id="B11">
  <title><p>{Controlling the opto-mechanics of a cantilever in an
  interferometer via cavity loss}</p></title>
  <aug>
    <au><snm>{Von Schmidsfeld}</snm><fnm>A.</fnm></au>
    <au><snm>Reichling</snm><fnm>M.</fnm></au>
  </aug>
  <source>Applied Physics Letters</source>
  <pubdate>2015</pubdate>
  <volume>107</volume>
  <issue>12</issue>
  <fpage>2</fpage>
  <lpage>-7</lpage>
</bibl>

<bibl id="B12">
  <title><p>{Fabry-P{\'{e}}rot resonator: spectral line shapes, generic and
  related Airy distributions, linewidths, finesses, and performance at low or
  frequency-dependent reflectivity}</p></title>
  <aug>
    <au><snm>Ismail</snm><fnm>N</fnm></au>
    <au><snm>Kores</snm><fnm>CC</fnm></au>
    <au><snm>Geskus</snm><fnm>D</fnm></au>
    <au><snm>Pollnau</snm><fnm>M</fnm></au>
  </aug>
  <source>Optics Express</source>
  <pubdate>2016</pubdate>
  <volume>24</volume>
  <issue>15</issue>
  <fpage>16366</fpage>
  <url>https://www.osapublishing.org/abstract.cfm?URI=oe-24-15-16366</url>
</bibl>

<bibl id="B13">
  <title><p>Calibration of rectangular atomic force microscope
  cantilevers</p></title>
  <aug>
    <au><snm>Sader</snm><fnm>JE</fnm></au>
    <au><snm>Chon</snm><fnm>JW</fnm></au>
    <au><snm>Mulvaney</snm><fnm>P</fnm></au>
  </aug>
  <source>Review of scientific instruments</source>
  <publisher>AIP</publisher>
  <pubdate>1999</pubdate>
  <volume>70</volume>
  <issue>10</issue>
  <fpage>3967</fpage>
  <lpage>-3969</lpage>
</bibl>

<bibl id="B14">
  <title><p>Single-electron charging using atomic force microscopy</p></title>
  <aug>
    <au><snm>Roy Gobeil</snm><fnm>A</fnm></au>
  </aug>
  <source>PhD thesis</source>
  <publisher>McGill University</publisher>
  <pubdate>2017</pubdate>
</bibl>

<bibl id="B15">
  <title><p>{Fully Quantized Electron Transfer Observed in a Single Redox
  Molecule at a Metal Interface}</p></title>
  <aug>
    <au><snm>Roy Gobeil</snm><fnm>A</fnm></au>
    <au><snm>Miyahara</snm><fnm>Y</fnm></au>
    <au><snm>Bevan</snm><fnm>KH</fnm></au>
    <au><snm>Grutter</snm><fnm>P</fnm></au>
  </aug>
  <source>Nano Letters</source>
  <pubdate>2019</pubdate>
  <volume>19</volume>
  <issue>9</issue>
  <fpage>6104</fpage>
  <lpage>-6108</lpage>
  <url>http://pubs.acs.org/doi/10.1021/acs.nanolett.9b02032
  https://pubs.acs.org/doi/10.1021/acs.nanolett.9b02032</url>
</bibl>

<bibl id="B16">
  <title><p>{Frequency modulation detection using high‐ Q cantilevers for
  enhanced force microscope sensitivity}</p></title>
  <aug>
    <au><snm>Albrecht</snm><fnm>T R</fnm></au>
    <au><snm>Gr{\"{u}}tter</snm><fnm>P.</fnm></au>
    <au><snm>Horne</snm><fnm>D</fnm></au>
    <au><snm>Rugar</snm><fnm>D</fnm></au>
  </aug>
  <source>Journal of Applied Physics</source>
  <publisher>AIP</publisher>
  <pubdate>1991</pubdate>
  <volume>69</volume>
  <issue>2</issue>
  <fpage>668</fpage>
  <lpage>-673</lpage>
  <url>http://aip.scitation.org/doi/10.1063/1.347347</url>
</bibl>

<bibl id="B17">
  <title><p>{The effective quality factor at low temperatures in dynamic force
  microscopes with Fabry–P{\'{e}}rot interferometer detection}</p></title>
  <aug>
    <au><snm>H{\"{o}}lscher</snm><fnm>H</fnm></au>
    <au><snm>Milde</snm><fnm>P</fnm></au>
    <au><snm>Zerweck</snm><fnm>U</fnm></au>
    <au><snm>Eng</snm><fnm>LM</fnm></au>
    <au><snm>Hoffmann</snm><fnm>R</fnm></au>
  </aug>
  <source>Applied Physics Letters</source>
  <pubdate>2009</pubdate>
  <volume>94</volume>
  <issue>22</issue>
  <fpage>223514</fpage>
  <url>http://aip.scitation.org/doi/10.1063/1.3149700</url>
</bibl>

<bibl id="B18">
  <title><p>{Quantification of antagonistic optomechanical forces in an
  interferometric detection system for dynamic force microscopy}</p></title>
  <aug>
    <au><snm>Tr{\"{o}}ger</snm><fnm>L</fnm></au>
    <au><snm>Reichling</snm><fnm>M</fnm></au>
  </aug>
  <source>Applied Physics Letters</source>
  <pubdate>2010</pubdate>
  <volume>97</volume>
  <issue>21</issue>
  <fpage>213105</fpage>
  <url>http://aip.scitation.org/doi/10.1063/1.3509412</url>
</bibl>

</refgrp>
} % end of \BMCxmlcomment

\end{backmatter}
\end{document}